\documentclass[a4paper]{jpconf}
\usepackage{graphicx}
\usepackage{amssymb}
\usepackage{amsmath}

\usepackage[
            colorlinks=true,
            linkcolor=red, 
            urlcolor=blue,
            citecolor=blue
            ]{hyperref}

\usepackage{cite}
\usepackage[font=small]{caption}
\newcommand{\blue}{\color{black}}

\newcommand{\black}{\color{black}}

\begin{document}
\title{Energy harvesting via co-locating  horizontal- and vertical-axis wind turbines}

\author{M. Hansen$^1$, P. Enevoldsen$^2$ and M. Abkar$^1$}

\address{$^1$Department of Engineering, Aarhus University, Denmark\\
$^2$Department of Business Development and Technology, Aarhus University, Denmark}

\ead{abkar@eng.au.dk}

\begin{abstract}
Co-locating horizontal- and vertical-axis wind turbines has been recently proposed as a possible approach to enhance the land-area power density of wind farms. 
In this work, we aim to study the benefits associated with such a co-location using large-eddy simulation (LES) and analytical wake models. In this regard, small-scale vertical-axis wind turbines (VAWTs) in triangular clusters are deployed within a finite-size wind farm consisting of horizontal-axis wind turbines (HAWTs). 
Wake flow within the wind farm and the effect of VAWTs on the overall wind-farm efficiency are investigated and quantified. 
The results show that the optimal deployment of small-scale VAWTs has a negligible impact on the performance of HAWT arrays while increasing the total power production. 
For the particular cases considered here, the power output of the co-located wind farm increases up to 21\% compared to the baseline case in which only the HAWTs are present.
Also, by comparing to the LES results, it is shown that the analytical framework proposed here is able to accurately predict the power production of wind farms including both HAWTs and VAWTs. 
Next, as a real-world application, potential benefits of deploying small-scale VAWTs inside the Horns Rev 1 wind farm are explored for various wind directions using the calibrated wake model. 
The results show potential for about an 18\% increase in the wind-farm power production, averaged over all wind directions, for a particular VAWT layout investigated in this study.  
\blue 
The levelized cost of energy (LCoE) for the co-located wind farm is also assessed. The simulations finds that meanwhile the installation of VAWTs increases the annual energy production 
of the wind farm, it also increases the LCoE, which is caused by a) lack of operational data, and b) a low technology readiness level
for VAWTs and floating foundations. \black
\end{abstract}

\section{Introduction}
\label{sec1}
Wind-turbine wakes require a relatively large distance to be fully recovered. Hence, when wind turbines are deployed in clusters, the performance of waked turbines significantly decreases compared to wind turbines in the free stream (see the review of Ref. \cite{Stevens2017}).   
One possible approach to mitigate the power defect due to the wake interaction is to install wind turbines as far as possible from one another \cite{Meyers2012}, which is further enforced by the development of rotor sizes throughout the past decades \cite{enevoldsen2019examining}.  However, this approach requires significant amounts of land which in practice is not always feasible  for several reasons ranging from the costs of aquiring the land to increased likelyhood of social opposition \cite{sivaram2018need,enevoldsen2018insights}. 

Historically, wind farms were assumed to consist of identical horizontal-axis wind turbines (HAWTs). Recent studies have proposed a paradigm shift in which size and type of turbines is also a decision variable in the farm-design process    \cite{vested2014wake,chamorro2014variable,Feng2017,Vasel-Be-Hagh2017,zhang2018large}. 
Feng and Shen \cite{Feng2017} investigated the benefits of wind farms consisting of HAWTs with multiple types using analytical wake models. They showed a lower energy cost for a wind farm with different sizes compared to a uniform-sized wind farm. 
Vasel-Be-Hagh and Archer \cite{Vasel-Be-Hagh2017} assessed the impact of hub-height optimization on wind-farm energy extraction. They found that a wind farm with variable hub heights can produce \textgreater $2$\% more energy annually compared to the wind farm with a uniform hub height. 
The benefits of vertically staggered wind farms by varying HAWT hub heights were also studied recently by Zhang et al. \cite{zhang2018large}. Using large-eddy simulation (LES), they showed that vertical staggering enhances the energy production of turbines in the entrance/developing region of the farm. However, this approach does not improve the power output in the fully developed regime.

A different approach is to fill the gap between large-scale HAWTs by deploying smaller vertical-axis wind turbines (VAWTs) \cite{dabiri2015new}. 
VAWTs are a class of turbines with rotational axes perpendicular to the free stream, and have received a great deal of attention in recent years (see for instance Refs. 
\cite{Dabiri2011,Shamsoddin2014,Bianchini2017,chatelain2017vortex,Rezaeiha2018b}, among others). 
VAWTs offer several advantages and opportunities over conventional HAWTs. In particular, they can produce power from any wind direction, thereby  obviate the need of any yaw control mechanism \cite{dabiri2015new}. 
They have also lower installation and maintenance costs as their drive train systems can be mounted close to the ground/sea surface \cite{Paraschivoiu2002}. 
Recently, Xie et al. \cite{Xie2016} performed LES of an infinite wind farm
(i.e., numerically subjected to periodic boundary conditions) 
consisting of co-located HAWTs and VAWTs. 
They showed that the small-scale VAWTs enhance the vertical momentum exchange within the farm leading to a significant increase (up to 32\%) in the total wind-farm power. 
Despite the promising findings reported in that study, the aerodynamic interaction of co-located HAWTs and VAWTs in a finite-size wind farm is unknown and has not been studied so far. Note that for a small number of wind turbines or at the leading edge of a large wind farm, the energy is mainly extracted from the incoming wind due the horizontal flux of kinetic energy \cite{Abkar2014}. 
Given the finite size of existing farms, investigating the impact of co-locating HAWTs and VAWTs on the overall performance of wind farms is valuable and is the central focus of this work. 

The present work aims at exploring wake flow and generated power from a finite-size wind farm consisting of co-located HAWTs and VAWTs using LESs and analytical wake models. Section \ref{chp:Modelling} provides a brief description of the LES and analytical frameworks for modeling wake flow through HAWTs and VAWTs. 
In Section \ref{chp:Results}, the impact of co-location on the wake flow and the total power production of the farm is examined, and the results are compared to the baseline case in which only HAWTs are present. 
Finally, the power enhancement by adding small-scale VAWTs to the Horns Rev 1 wind farm, as an existing wind farm, is investigated for various wind directions.
In Section \ref{chp:Conclusion}, a summary and concluding remarks are given.

\section{Modelling}\label{chp:Modelling}
\subsection{Large-eddy simulation framework}
The previously-validated LES framework presented here (see for instance Refs. \cite{Wu2015,Abkar2015a,Abkar2017a,Abkar2018}) 
solves the filtered continuity and Navier–Stokes equations for incompressible turbulent flow,  
\begin{equation}
\frac{\partial \tilde{u}_i}{\partial x_i}=0,  
\frac{\partial\tilde{u}_i}{\partial t}+
\frac{\partial (\tilde{u}_i \tilde{u}_j)}{\partial x_j} =
-\frac{1}{\rho_o} \frac{\partial \tilde{p}}{\partial x_i}
-\frac{\partial \tau_{ij}}{\partial x_j} 
-\frac{f_i}{\rho_o},
\end{equation}
where ${\tilde{u}}_i$ and ${\tilde{p}}$ are the filtered velocity and pressure fields, respectively.  
$x_i$ indicates the Cartesian coordinates. 
$t$ is time. $\rho_o$ is the fluid density. 
$\tau_{ij}=\widetilde{u_i u_j}- \tilde{u}_i \tilde{u}_j$ denotes the kinematic subfilter stress tensor. 
$f_i$ is a body force and accounts for the effect of wind turbines on the flow.
The code employs a pseudo-spectral discretization in horizontal directions, and a central finite difference method in the vertical direction. The second-order Adam-Bashforth scheme is applied for time advancement.
The molecular viscous forces are neglected away from the wall, hence the flow is at nominally infinite Reynolds number.  
The subfilter turbulent motions are modeled via the scale-dependent Lagrangian dynamic approach \cite{Bou-Zeid2005}. 
The actuator-disk model with rotation \cite{Wu2015} and the actuator swept-surface model \cite{Shamsoddin2014} are respectively used to parameterize the forces induced by HAWTs and VAWTs.
Through these approaches, the aerodynamic forces on the rotors are determined
using the blade airfoil geometry, the relative wind velocity, and the lift-drag force characteristics of the blades. 
The inflow condition is generated through a precursor method and by simulating a turbulent flow over a rough surface. The computational domain size is $4800\text{m} \times 1200\text{m}\times 456\text{m} $ in the 
streamwise ($x$), lateral ($y$) and wall-normal ($z$) directions, 
respectively, and it is broken uniformly into $480\times 240\times 96$ grid points. 
An imposed uniform pressure gradient drives the boundary-layer flow in the streamwise direction. The effective ground roughness ($z_o$) is $0.05$m, and the wall shear stress is specified following the the equilibrium wall model. 
Further details of the solver can be found in Ref. \cite{Yang2018}. 

\begin{figure}
\includegraphics[width=19pc]{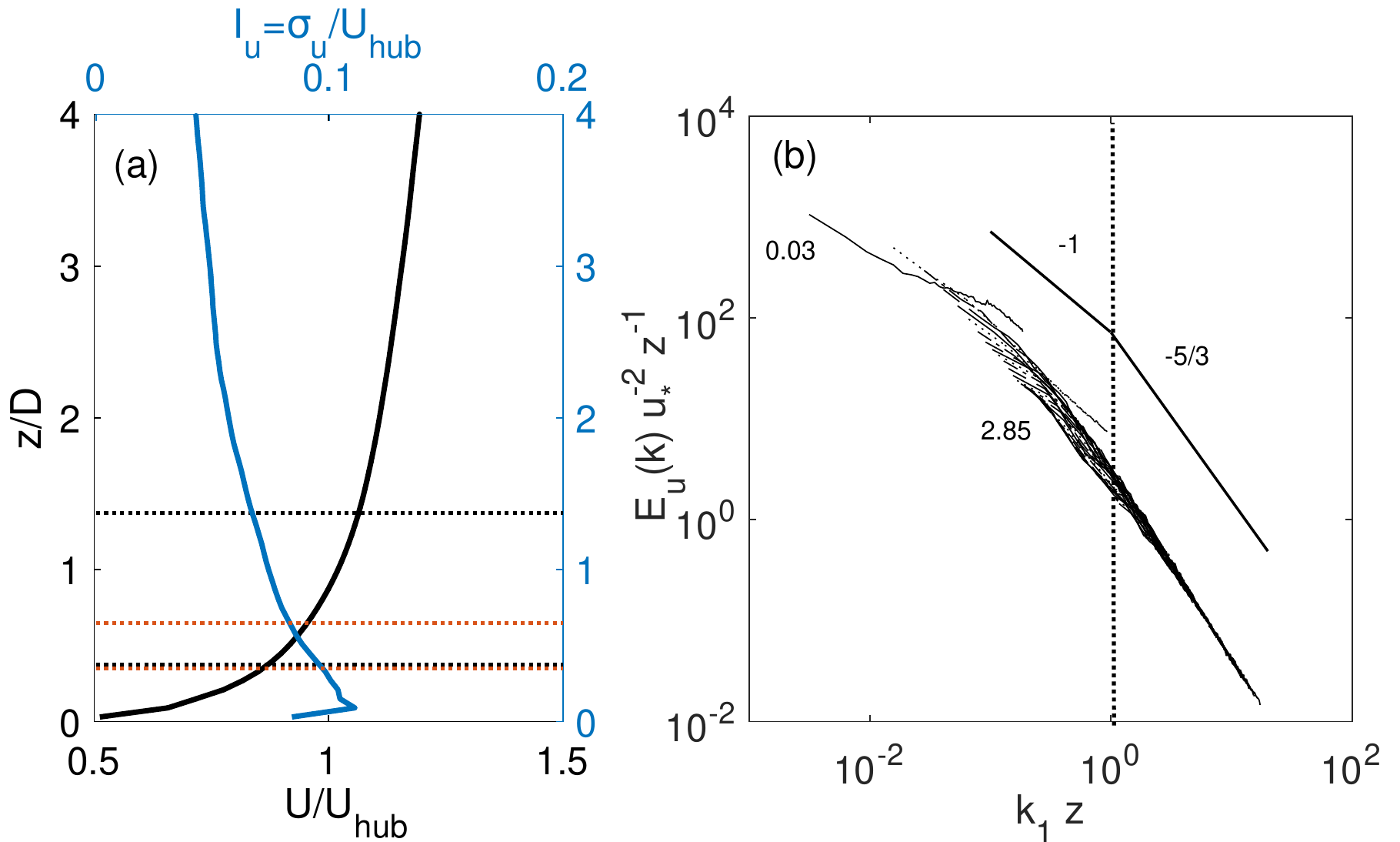}\hspace{2pc}%
\begin{minipage}[b]{16pc}\caption{\label{fig:inflow}(a) Vertical profiles of the normalized mean streamwise velocity $U/U_{hub}$ and the turbulence level $I_u=\sigma_u/U_{hub}$ for the incoming free stream. Horizontal black and red dot-lines show the HAWT and VAWT extents, respectively. (b) Normalized resolved streamwise velocity spectra. Here, $u_*$ is the friction velocity, and the normalized height $z/D$ increases from 0.03 to 2.85.}
\end{minipage}
\end{figure}

The main characteristics of the incoming wind including the mean velocity, turbulence intensity and power spectra of the streamwise velocity component are shown in Fig. \ref{fig:inflow}. 
The free-stream mean velocity at the HAWT hub-height $U_{hub}$ is about $7.3$m/s. The streamwise turbulence level $I_u=\sigma_u/U_{hub}$ at the same height is about $7.4$\%, where $\sigma_u$ is the standard deviation of the streamwise velocity. 
Figure \ref{fig:inflow}b illustrates the normalized spectra of the simulated streamwise velocity field obtained from the precursor simulation.
As can be seen in this figure, the normalized power spectra depict the expected collapse for the small resolved scales ($k_1z>1$) and follow the theoretical scaling in the inertial subrange with a slope of $-5/3$.  
In the wake-flow simulation, a fringe zone is implemented to adopt the flow from the wake state downstream to that of a fully turbulent boundary-layer inflow condition \cite{stevens2014concurrent,Abkar2015a}. 

\begin{figure} [h]
\includegraphics[width=15pc]{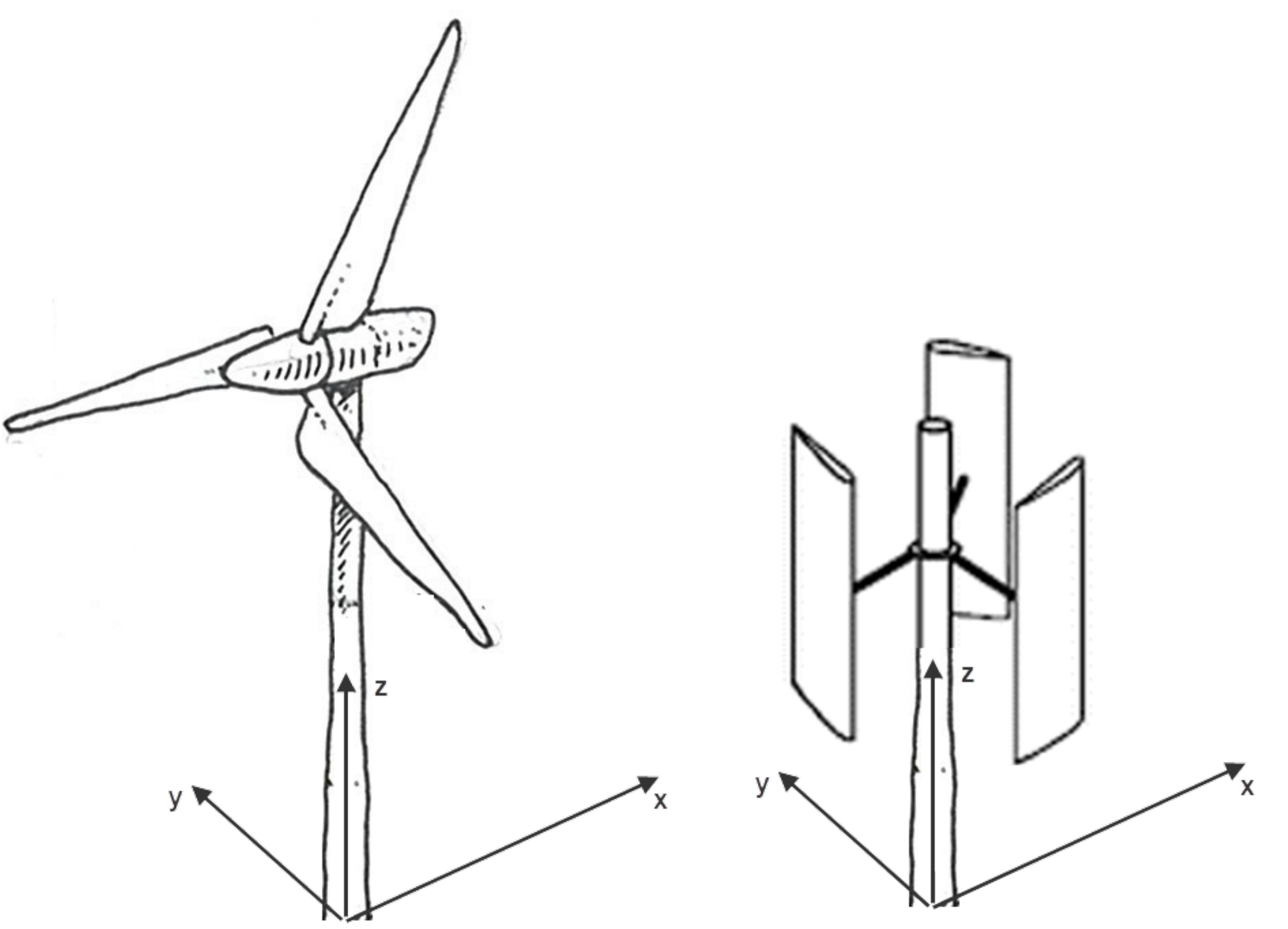}\hspace{2pc}%
\begin{minipage}[b]{20pc}\caption{\label{fig:schematic}A schematic of a HAWT (left), and a three- and straight-bladed VAWT (right).}
\end{minipage}
\end{figure}

The HAWTs used in the simulations are Vestas V80 turbines with the rotor diameter ($D$) of $80$m and the hub height ($z_h$) of $70$m. Details of the Vestas V80 wind turbine such as distributions of twist angle and chord length along the blades, and lift-drag coefficients as a function of angle of attack can be found in Ref. \cite{Wu2015}. 
The VAWTs immersed in the flow are 200kW T1-turbines, which are the three- and straight-bladed VAWTs with the rotor diameter ($D_v$) of $26$m, the blade span ($H_v$) of $24$m, and the equator height ($z_{hv}$) of $40$m. The blades of VAWTs consist of the standard NACA $0018$ airfoil with the chord length of $0.75$m, and the turbines operate at the nominal tip-speed ratio of $3.8$ \cite{Abkar2017a}. The thrust coefficients of the HAWTs and VAWTs are respectively around 0.8 and 0.64.   
A schematic of the HAWT and VAWT is shown in Fig. \ref{fig:schematic}. 

\subsection{Analytical framework} 
The reduction of the mean wind velocity inside the wake is described by the normalized velocity deficit as ${\Delta U}/ U_{\infty}=({U_{\infty} -U_w})/{U_{\infty}}$, 
where $U_{\infty}$ is the turbine inflow velocity, and $U_w$ is the wind velocity at a given position inside a wake.
In this study, we use the recently-introduced Gaussian wake model for both HAWTs \cite{Bastankhah2014,Abkar2018a} and VAWTs \cite{Abkar2017a,Abkar2019}. 
The Gaussian wake model describes the wake velocity deficit as 

\begin{equation}\label{eq:gaussianmodel}
    \frac{\Delta U}{U_{\infty}} = C(x) \times \exp \left( -\frac{1}{2} \left[ \left( \frac{ y }{ \sigma_y}\right)^2+\left( \frac{ z-z_h }{ \sigma_z}\right)^2 \right]\right) ,
\end{equation}
where $C$ is the maximum velocity deficit, and $\sigma_y$ and  $\sigma_z$ are respectively the spanwise and wall-normal standard deviations of velocity deficit distribution. 
The maximum velocity deficit in the wake is given by
$C = 1-\sqrt{1-{C_t}/[{2\pi \left( \sigma_y \sigma_z / A_p \right)]}}$. 
Here, the turbine thrust coefficient is denoted by $C_t$. $A_p$ is the turbine projected area, and is equal to $\pi D^2/4$ and $D_vH_v$ for HAWTs and VAWTs, respectively. 
As shown in the earlier studies \cite{Bastankhah2014,Abkar2019}, 
for wind turbines in the turbulent free stream, 
a linear growth of the wake with downwind distance can be assumed, $\sigma_y = k^* x + \epsilon L_y$ and $\sigma_z = k^* x + \epsilon L_z$, 
where $k^*$ denotes the wake expansion rate, and the characteristic turbine dimensions in the $y$ and $z$ directions are respectively labeled as $L_y$ and $L_z$. In particular, $L_y = L_z = D$  
for HAWTs, and $L_y = D_v$ and $L_z = H_v$ for VAWTs \cite{Abkar2019}. 
The wake expansion rate $k^*$ is estimated based on the empirical formula suggested in Ref. \cite{CarbajoFuertes2018} as $k^* = 0.35I_u$. 
$\epsilon$ in the equation above characterizes the wake standard deviation at the rotor, and it is defined as $\epsilon = 0.25 \sqrt{\beta}$, where $\beta = 0.5 (1+\sqrt{1-C_t})/\sqrt{1-C_t}$ \cite{Bastankhah2014}. 
Note that in the very far-wake region, where $k^* x \gg \epsilon L_{y,z}$, the velocity deficit distribution by definition asymptotes to a circular shape from an elliptic shape as $\sigma_y/\sigma_z \to 1$.   
To account for multiple wake interactions within a wind farm, rotor wakes can be superposed in either linear or nonlinear manner \cite{Lissaman1979,Voutsinas1990}. 
Based upon the linear superposition , the wake velocity within the farm can be estimated as $U_i = U_{\infty} - {\sum_k \left( U_k - U_{ki} \right)}$   
where $U_i$ is the velocity at turbine $i$, $U_k$ is the incoming velocity at turbine $k$, and $U_{ki}$ is the wake velocity caused by turbine $k$ at downstream turbine $i$. 
The velocity field emerged from multiple wake interactions can be alternatively modeled using the nonlinear superposition method  as $U_i = U_{\infty} - \sqrt{\sum_k \left( U_k - U_{ki} \right)^2}$. 
In the following section, the performance of the above-mentioned superposition methods is assessed using the LES data. 
The generated power by turbine $i$ is determined as $P_i = 0.5\rho_o C_p A_p U_i^3$, where $C_p$ is the power coefficient of the turbine.

\blue
It should be mentioned that the wake model for HAWTs has been validated using wind-tunnel measurements and LES data \cite{Bastankhah2014}, as well as field experiments of wind-turbine wakes \cite{CarbajoFuertes2018}. The wake model for VAWTs has been recently assessed using the LES data as well as field measurements of VAWT wakes\cite{Abkar2019}. The derivation of the analytical wake model for both HAWTs and VAWTs is based on assuming a two-dimensional Gaussian shape for the wake velocity deficit, and assuming a linear growth rate for the wake expansion downstream of the turbine. Hence, the accuracy of the analytical models is sensitive to the validity of the aforementioned assumptions. Besides, selecting the magnitude of the wake expansion rate $k^*$, as a tuning parameter, together with the employed superposition model can cause uncertainties in the prediction of wake flows and power output in wind farms.
\black


\subsection{Wind-farm layout}
\label{layout}
\begin{figure}
\includegraphics[width=22pc]{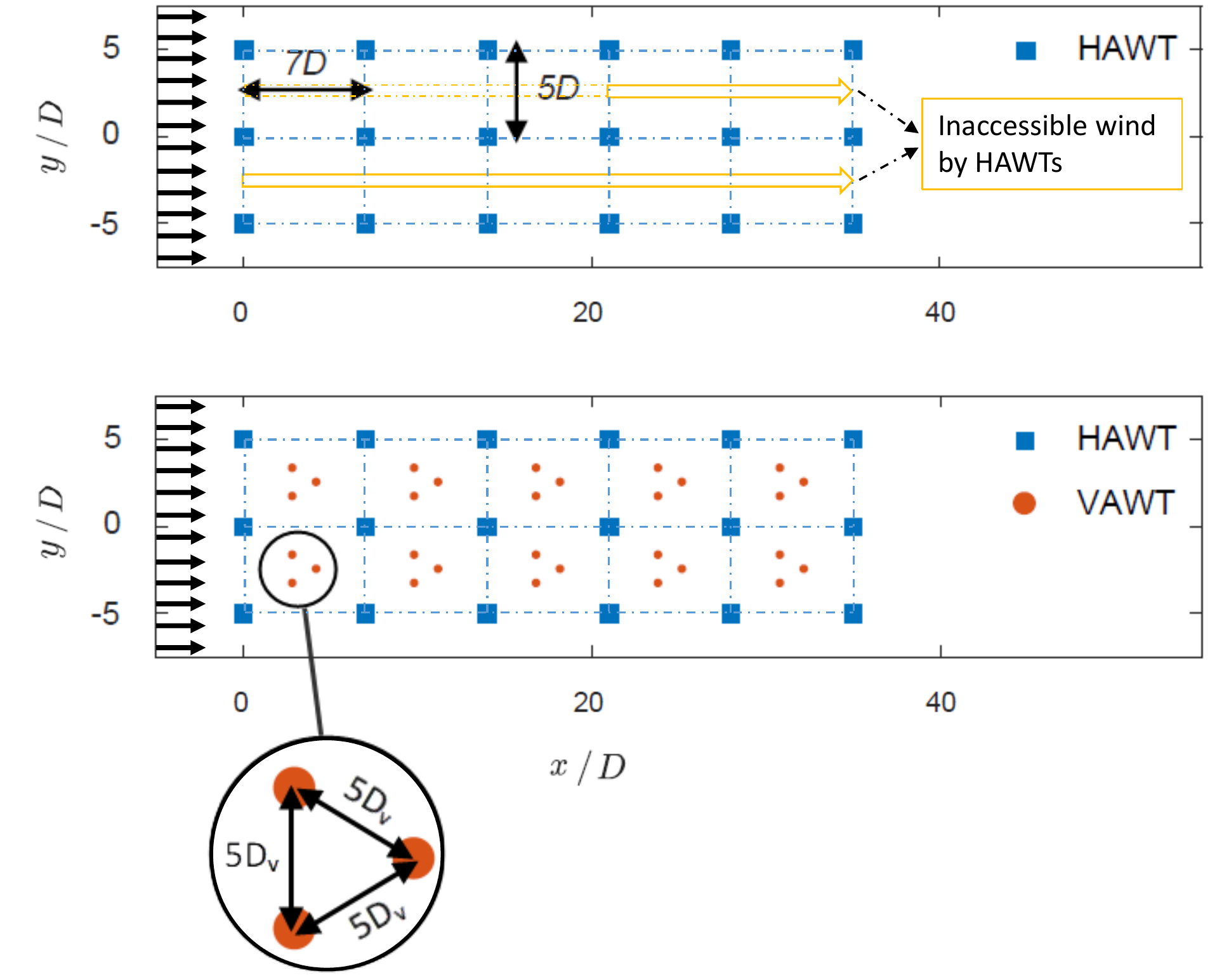}\hspace{2pc}%
\begin{minipage}[b]{14pc}\caption{\label{fig:layoutcases}Wind-farm layout: Case (0): baseline (top) and Case (1): VAWT clusters between HAWTs (bottom).}
\end{minipage}
\end{figure}

Figure \ref{fig:layoutcases} illustrates the schematic of the wind-farm layout for the two cases considered in this study.  
Case (0) is the baseline case with no VAWT. 
The HAWTs are arranged in six columns and three rows in the streamwise and lateral directions, respectively. 
The distance among the HAWTs in the streamwise and lateral directions is $7D$ and $5D$, respectively. 
As mentioned before, deployment of wind turbines as far apart is required to mitigate the wake loss in wind farms. However, the available energy passing through the gap between large HAWTs is not accessible by them. In order to increase the land-area power density of wind farms, the gap between large HAWTs can be filled by the smaller VAWTs \cite{sivaram2018need}. 
There are different ways to fill the gap between HAWTs using VAWTs, and several design parameters such as geometry, number, and exact location of VAWTs can be essentially optimized in order to maximize the benefits of wind-farm co-location. In this study, we consider one particular case to show the potential benefits of this approach in improving the power production of an existing wind farm. 
Case (1) represents a layout in which VAWTs are installed in triangular clusters in the free space among HAWTs. 
Here, the center of VAWT cluster is placed between each row and column of HAWTs in order to minimize the interaction between them. 
Each VAWT cluster consists of three turbines with center-to-center distance among them of about $s_vD_v = 5D_v$. 
The selected configuration for the VAWT clusters is motivated by the recent study by Hezaveh et al. \cite{Hezaveh2018} who showed that the performance of VAWTs can be increased using synergetic clustering. In particular, by optimizing the VAWT numbers and the distance among them, they showed that such a configuration results in higher power production over a wide range of wind direction. 
As can be visually acknowledged in Fig. \ref{fig:layoutcases}, the distance between VAWT clusters in the streamwise and lateral direction is $7D\simeq21.5D_v$ and $5D\simeq15.4D_v$, respectively, and the first column of VAWT clusters is placed $3.5D$ downstream of the first HAWT column.   

\section{Results}\label{chp:Results}
Here, we present the results obtained from the LES and the analytical wake model. In particular, we focus on analyzing the mean wake flow through the wind farm as well as the generated power by wind turbines for the two scenarios described in Section \ref{layout}.

\subsection{Case (0): HAWT farm}
\begin{figure}
\includegraphics[width=22.5pc]{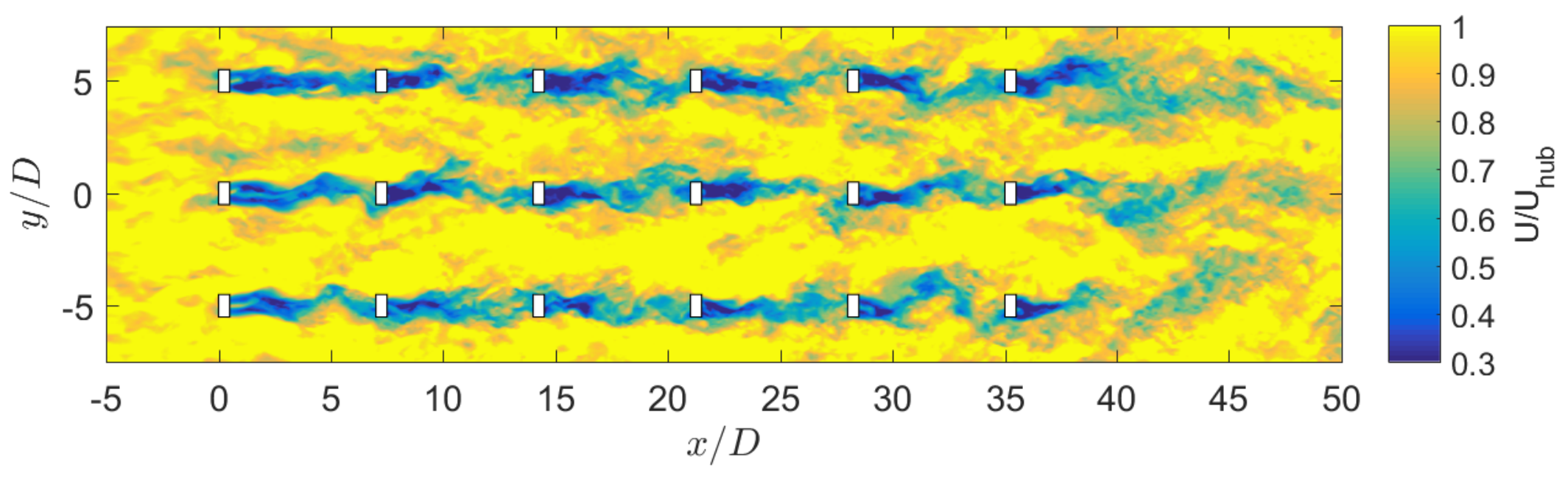}\\
\includegraphics[width=22.5pc]{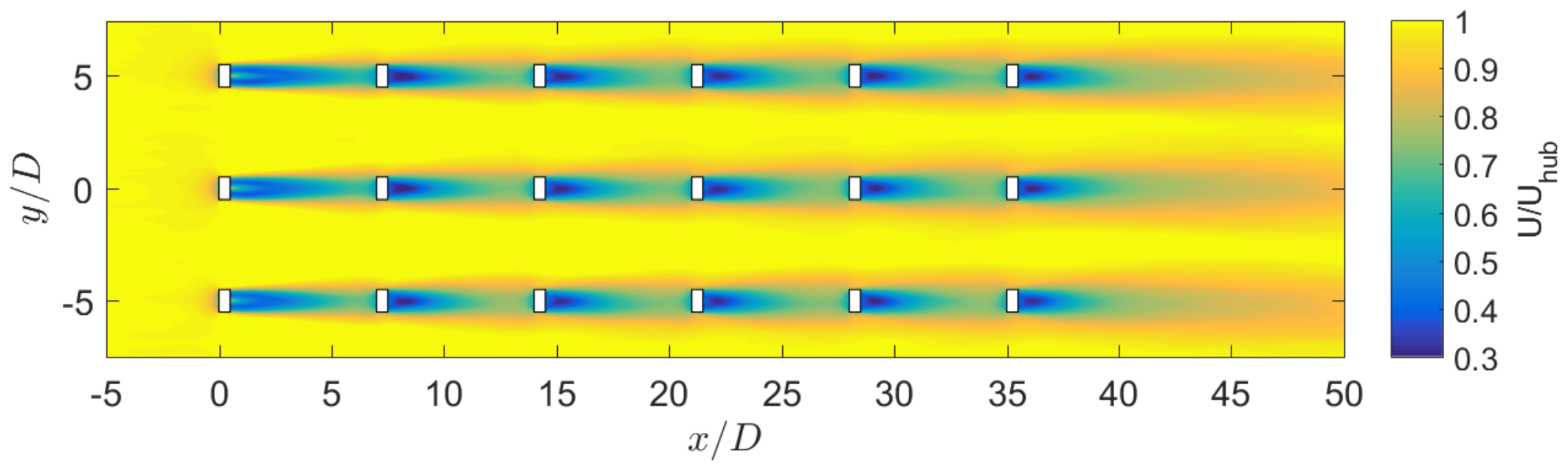}\hspace{2.pc}%
\begin{minipage}[b]{13pc}\caption{\label{fig:case0contourzh}
Instantaneous (top) and time-averaged (bottom) streamwise velocity field at HAWT hub height obtained from LES for Case (0).}
\end{minipage}
\end{figure}

\begin{figure}
\includegraphics[width=22pc]{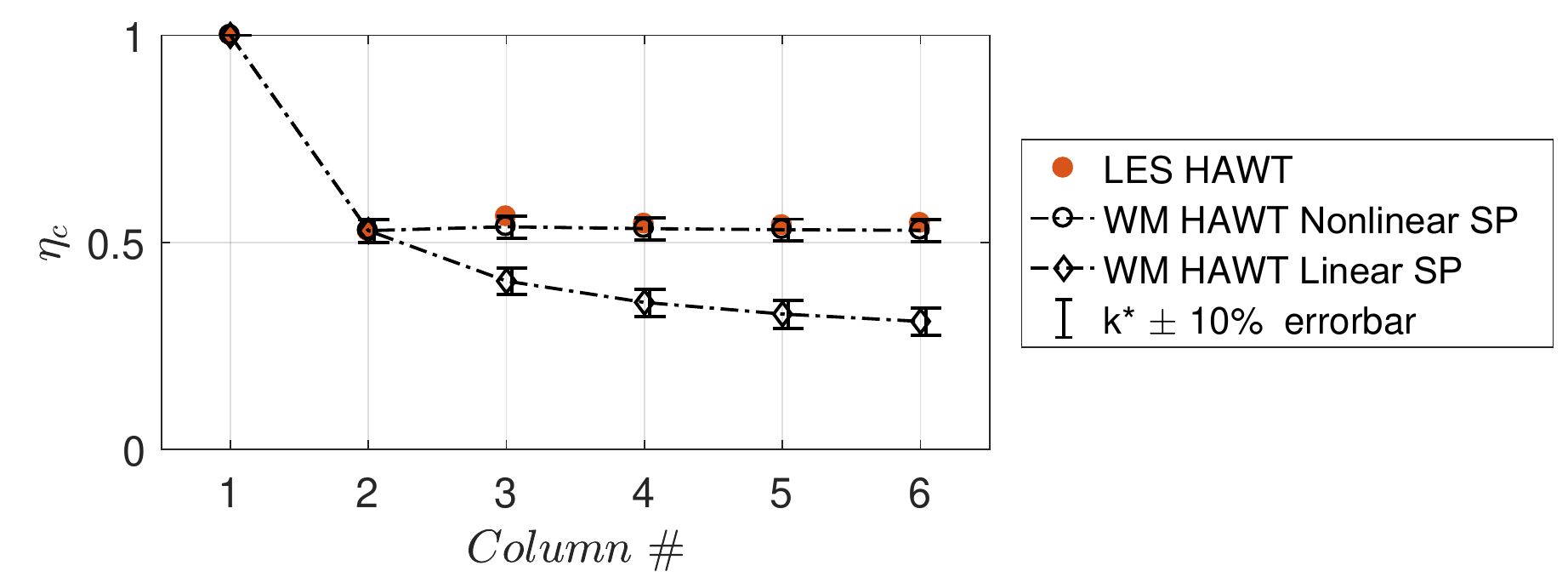}
\hspace{2pc}
\begin{minipage}[b]{13pc}\caption{\label{fig:case0eta}Power efficiency as a function of turbine columns for Case (0) obtained from LES and wake model (WM) with different superposition (SP) methods.}
\end{minipage}
\end{figure}

Figure \ref{fig:case0contourzh} illustrates contour plots of the normalized instantaneous and time-averaged streamwise velocity at the HAWT hub height obtained from LES. As expected, the waked turbines encounter a reduced wind velocity and, consequently, produce less power compared to the turbines in the first column.  
Figure \ref{fig:case0eta} represents the power efficiency as a function of turbine columns. 
Here, the power efficiency of each column is calculated as $\eta_c = {P_c}/{P_{c,free}}$,  
where $P_c$ is the power output of all turbines in each column, and $P_{c,free}$ is the power output of all HAWTs in the first column operating in the free stream.
The first column has all the turbines in the free stream and, hence, $\eta_c = 1$. Due to the wake effect, the efficiency drops by around $50 \%$ for turbines in the downstream columns. 
In this figure, the predictions obtained from the analytical wake model with linear and nonlinear superposition methods are also provided for comparison. As can be seen, the linear addition of the wake velocity defects leads to an underestimation of the power output for the downwind wind turbines. On the other hand, there is a good agreement between LES and the analytical model with nonlinear wake superposition. 
These results are consistent with previous studies that show the nonlinear superposition method can provide a more accurate prediction for the power production of wind turbines subjected to multiple wake interactions (see for instance the review of Refs. \cite{goccmen2016wind,Archer2018}, among others).   
In the analytical wake model, we also investigate the uncertainty in estimating the wake growth rate, $k^*$, as it is the only empirically-tuned parameter in the wake model. Here, we present the results with 10\% uncertainty in the estimation of this parameter. It is found that 10\% uncertainty in the wake growth rate leads to about {2.3\%} uncertainty in the power prediction of waked wind turbines. 
Since the analytical model with the nonlinear wake superposition provides a relatively accurate prediction for the power output in the baseline case, this method is used in the rest of the paper.

\subsection{Case (1): Co-located VAWT clusters and HAWTs}

\begin{figure}
\includegraphics[width=22.5pc]{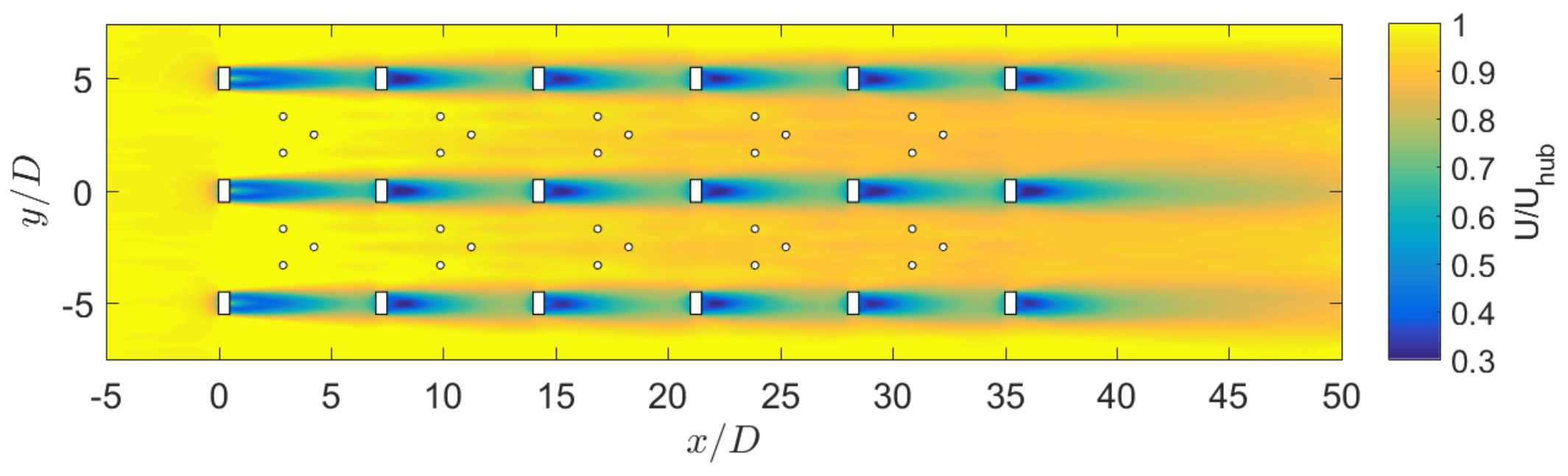}\\
\includegraphics[width=22.5pc]{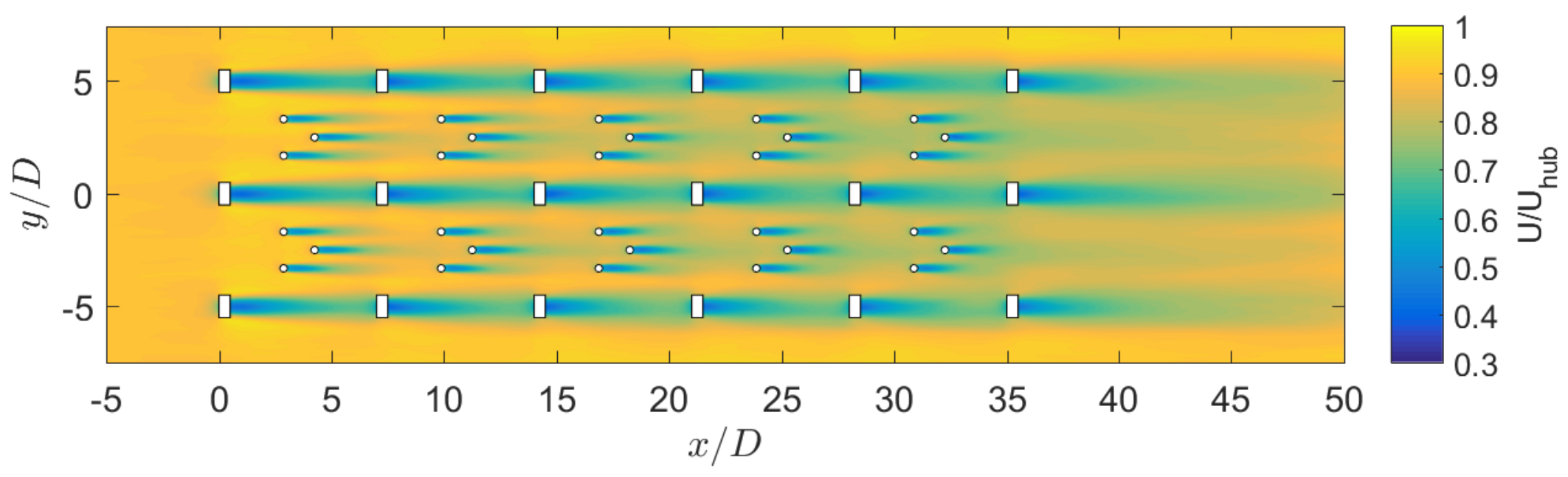}\hspace{2.pc}%
\begin{minipage}[b]{13pc}\caption{\label{fig:case4contourzv} Time-averaged streamwise velocity field at HAWT (top) and VAWT (bottom) hub heights obtained from LES for Case (1).}
\end{minipage}
\end{figure}

In this case, HAWTs are kept in the same position as Case (0), and VAWT clusters are placed in the gap among them. Figure \ref{fig:case4contourzv} show the contours of the normalized mean flow at HAWT and VAWT hub heights, respectively, obtained from LES. At the HAWT hub height, VAWT wakes appears far downstream of the farm, since the wake behind VAWTs needs to grow sufficiently in the vertical direction before being visible at the HAWT hub height. 
It can be also realized from these figures that the wakes of VAWT clusters are almost fully recovered before the next column since the relative distance between the two columns of VAWT clusters is relatively large (around $21.5 D_v$). 

\begin{figure}
\includegraphics[width=22pc]{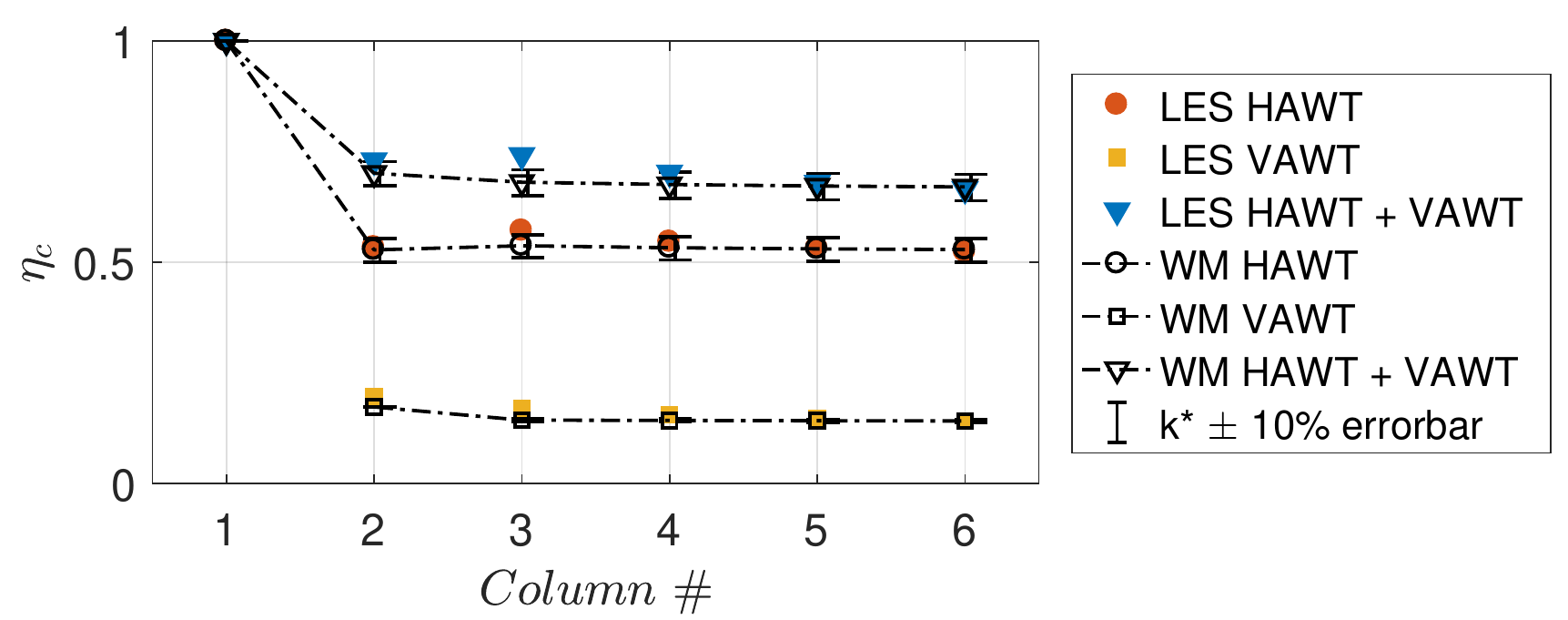}
\hspace{2pc}
\begin{minipage}[b]{13pc}\caption{\label{fig:case1eta}Power efficiency as a function of turbine columns for Case (1) obtained from LES and wake model (WM).}
\end{minipage}
\end{figure}

The average turbine power efficiency for the turbines in each column is plotted in Fig. \ref{fig:case1eta}. 
The total power efficiency of the co-located wind farm is also plotted in this figure to illustrate the gain potential due to the presence of VAWTs. 
In order to better quantify the effect of VAWTs on the performance of HAWTs and on the total power output of the farm, we define the gain/loss factor as follows.
The gain/loss factor in the power production of HAWTs due to the presence of VAWTs is defined as $\zeta_{HAWT,n} = {\sum P_{HAWT,n}}/{\sum P_{0}}-1$, 
where $\sum P_{HAWT,n}$ is the total power of HAWTs in Case (n) and $\sum P_{0}$ is the total power of HAWTs in the baseline case (i.e., Case 0).
We also define the gain factor for VAWTs as $\zeta_{VAWT,n} = {\sum P_{VAWT,n}}/{\sum P_{0}}$, 
where $\sum P_{VAWT,n}$ is the total power of VAWTs in Case (n).
Then, the net gain/loss factor is defined as $\zeta_{net,n} =\zeta_{HAWT,n} + \zeta_{VAWT,n}$. 
Note that, based on the definitions above, the gain/loss factor for Case (0) is zero. 
The gain/loss factor for the co-located wind farm is provided in Table \ref{tab:zeta}. 
As can be noticed, the VAWTs have a relatively small impact on the performance of HAWTs by decreasing the HAWT farm efficiency by $0.7\%$. 
The slight reduction of the HAWT farm efficiency is related to the presence of VAWTs and their wake interactions with the HAWTs as shown in Fig. \ref{fig:case4contourzv}. 
However, these interactions are relatively small. Besides, due to the additional power generated by the VAWTs, the total power output is increased by $21.1\%$ compared to the baseline case. 
Note that optimizing the design and placement of VAWTs can further reduce their negative impacts on HAWTs, and thus it can be addressed in future works. 
Also, similar to the previous results, a relatively good agreement between the LES data and the wake model is observed in predicting power efficiency for both the HAWTs and VAWTs. 

In this study, in order to assess the effect of HAWTs on VAWTs, we also considered another case in which only VAWTs are present (not shown here). We found that the power output of VAWT only farm is slightly less (about 8\%) than the VAWTs in HAWT+VAWT case. The reason is mainly related to the fact that in aligned HAWT farms, there is a high-speed “channels” between the turbine rows. In particular, the wind speed slightly increase in the channels among the HAWT rows due to the blockage induced by the turbines. In the co-located wind farm, since the VAWTs are located between the HAWTs in the high-speed channels, the power output from the VAWTs is slightly more than the one from the VAWT only farm. 

\begin{table}
    \centering
\def~{\hphantom{0}}    
    \caption{Gain/loss factor for Case (1) obtained from LES. 
    }
    \label{tab:zeta}
    \begin{tabular}{l c c c}
 \hline\noalign{\smallskip}    
$I_u$ & $\zeta_{HAWT}$     &  $\zeta_{VAWT}$  & $\zeta_{net}$               \\
 \noalign{\smallskip}\hline\noalign{\smallskip} 
$7.4\%$  &  $-0.7\%$ & $21.8 \%$ & $21.1 \%$  \\
 \hline
    \end{tabular}
\end{table}

\subsection{Horns Rev 1 wind-farm power enhancement}
\begin{figure}
	\centering
	\hspace*{-0.5cm}
	\includegraphics[width=.425\textwidth]{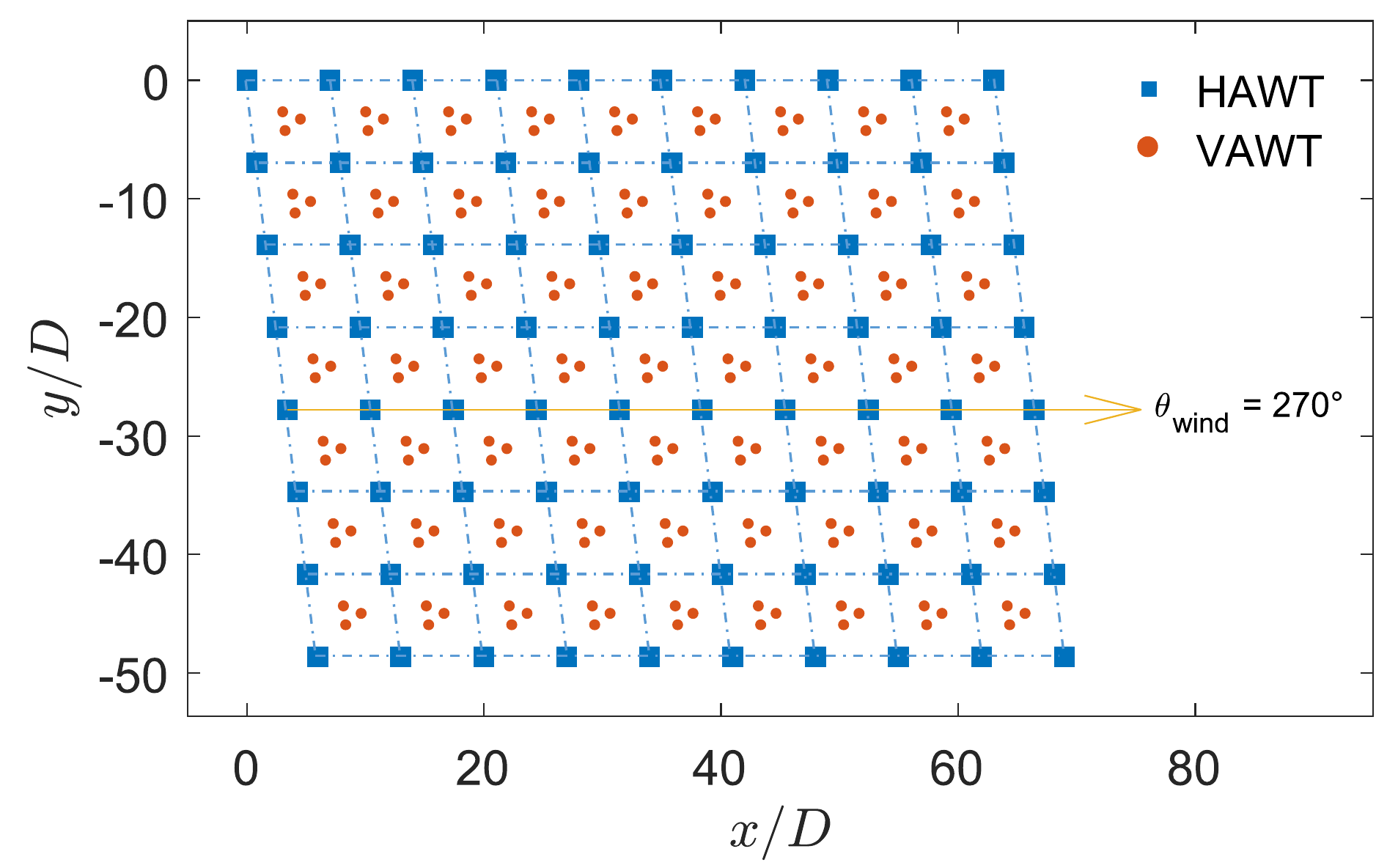}
	\includegraphics[width=.48\textwidth]{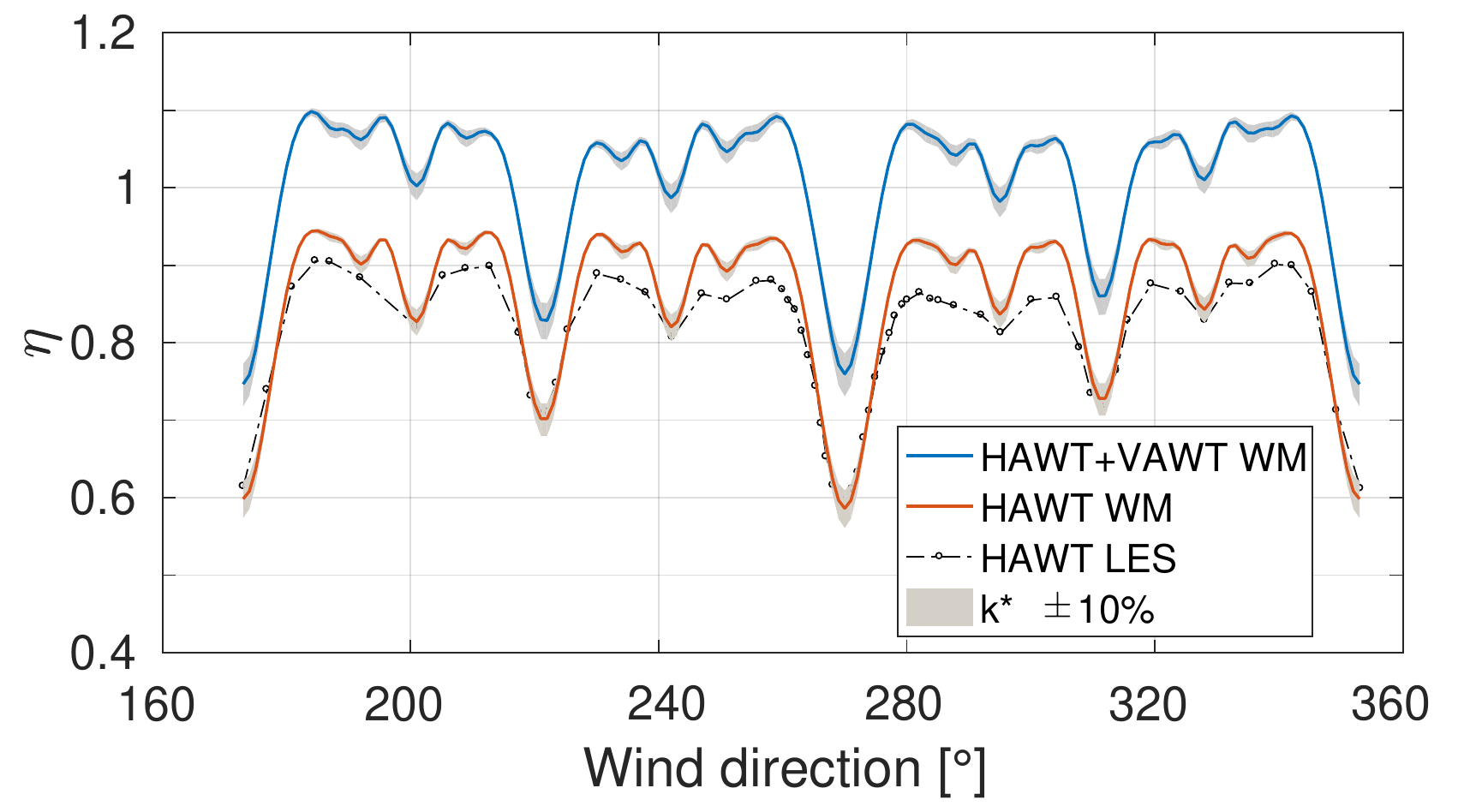}
	\caption{Left: Layout of the Horns Rev 1 wind farm. 
	The HAWTs and VAWTs are respectively shown with blue squares and red circles.
	Right: Power efficiency distribution of the Horns Rev 1 wind farm with VAWT clusters. 
	The ambient turbulence level is $7.7\%$. LES data were digitally extracted from Ref. \cite{Niayifar2016}. WM denotes the analytical wake model. 
	}
	\label{fig:HornsRev_layout}
\end{figure}

As a real-world application, the benefits of co-locating HAWTs and VAWTs are investigated in the Horns Rev 1 wind farm under different wind directions. To do so, the analytical wake model, calibrated in the previous section, is employed.   
The Horns Rev 1 wind farm consists of 80 Vestas V80 HAWTs in 8 rows by 10 columns grid with a minimum distance among two turbines ($s$) of 7 HAWT rotor diameter. Columns are turned  $7^{\circ}$ from the North-South axis. A schematic of the wind farm is shown in Fig. \ref{fig:HornsRev_layout}. The type of turbines is the same as the ones used in previous cases. 
The incoming wind velocity at the HAWT hub height is $8$m/s. To assess the effect of atmospheric turbulence on the results, three different values for the ambient turbulence intensity are considered as $I_u=5\%$, $7.7\%$ and $15\%$.
VAWTs are deployed based on the same layout concept as Case (1), i.e., VAWT triangular clusters in the gap between the HAWTs. 
In order to provide a more complete picture of the effect of VAWTs on the HAWT array, wind-farm efficiency in all wind directions is investigated. Due to the symmetry of the wind-farm layout, it is sufficient to investigate wind directions ranging over $180^{\circ}$. 
Here, we define the efficiency of the entire wind farm as $\eta = {\sum P}/{\sum P_{free}}$, where $\sum P$ is the sum of power output from all wind turbines, and $\sum P_{free}$ is the total power output from HAWTs calculated as if they were placed in the free stream. 

Figure \ref{fig:HornsRev_layout} shows the wind-farm power efficiency, $\eta$, for wind directions ranging from  $173^{\circ}-353^{\circ}$  obtained from wake model with 10$\%$ uncertainty in the wake growth rate. 
The wind-farm power output obtained from LES in the absence of VAWTs \cite{Niayifar2016} is also included for comparison. 
As can be seen, a relatively good agreement between the LES and the wake model is observed, especially for wind directions with lower power efficiency under full-wake conditions. 
In other wind directions, under partial-wake regimes, the wake model overestimates the power efficiency of the farm. 
It should be noted that, although there are some discrepancies between the LES and wake model, in this study, we aim to investigate the relative improvement of wind-farm power production due to the presence of VAWTs. 
Adding the VAWTs to the wind farm shows a similar trend in power efficiency as the one discussed earlier. 
The effect of small-scale VAWTs on the HAWTs is relatively small, while they produce additional wind power. It can be also seen that, for all wind directions, the efficiency of the co-located wind farm is higher than the baseline case.  
To better quantify the impact of VAWTs on the HAWT farm, the gain/loss factor under different ambient turbulence levels is provided in Table \ref{tab:zeta_HR}. 
For the particular wind-farm layout considered here, the power production of the farm (averaged over all wind directions) increases by $16.5\%$, $17.3\%$ and $18.2\%$ for the ambient turbulence intensity of $5\%$, $7.7\%$ and $15\%$, respectively. 
It is interesting to observe that the ambient turbulence level has a relatively small effect on power enhancement due to the turbine co-location. 
In addition, it is found that $10 \%$ uncertainty in the wake growth rate yields about $1\%$ uncertainty in estimating the total power production averaged over all wind directions which is much smaller than the gained power due to the presence of VAWTs. 
This result reveals the potential to enhance the wind-power density of wind farms by co-locating VAWTs and HAWTs. 

\begin{table}
    \centering
\def~{\hphantom{0}}    
    \caption{Gain/loss factor for the Horns Rev 1 case averaged over all wind directions. 
    }
    \label{tab:zeta_HR}
    \begin{tabular}{l c c c}
 \hline\noalign{\smallskip}    
$I_u$ & $\zeta_{HAWT}$     &  $\zeta_{VAWT}$  & $\zeta_{net}$               \\
 \noalign{\smallskip}\hline\noalign{\smallskip} 
5\%   & $-1.4 \%$  & $ 17.9 \%$ & $16.5\%$  \\ 
7.7\% & $-1.0 \%$  & $ 18.3 \%$ & $17.3\%$   \\
15\%  & $-0.6 \%$  & $ 18.8 \%$ & $18.2\%$  \\
 \hline
    \end{tabular}
\end{table}

\subsection{The levelized cost of energy}
The levelized cost of energy (LCoE) is measured as \$/MWh and is an important parameter for benchmarking the competitiveness of any energy technology \cite{aldersey2019levelised}. The LCoE is estimated following equation, 
\begin{equation}\label{LCoE}
LCoE=\frac{\sum_{t=1}^{n} \left[CAPEX_t+\sum_{t=1}^{n}\left(\frac{OPEX_t}{(1+r)^t}+\frac{DECEX}{(1+r)^t}\right)\right]}{\sum_{t=1}^{n} \frac{E_t}{(1+r)^t}},
\end{equation}
where CAPEX describes the capital expenditures with hardware as the greatest cost, OPEX describes the operational expenditures such as service and maintenance, and finally DECEX describes the expenditures related to decommissioning of the wind turbine. 

\blue
The expected LCoE for European offshore wind in 2028 is 51 \$/MWh for fixed-bottom foundations \cite{Mackenzie2019}. 
For the studied wind farm, Horns Rev 1, the mean annual energy production (AEP) is 577.64 GWh/year estimated over the past eleven years (2009-2019)\footnote{Data obtained from https://ens.dk/service/statistik-data-noegletal-og-kort/data-oversigt-over-energisektoren.}, which have been used to estimate the potential impact of combining VAWT and HAWT in Table \ref{tab:AEP}. 
Having estimated the corresponding energy outputs, Equation \ref{LCoE} can be applied to determine the impact on LCoE of an offshore wind farm with and without the installation of VAWTs. Table \ref{tab:LCoE} provides an overview of such.

\begin{table}
    \centering
\def~{\hphantom{0}}    
    \caption{The mean AEP (GWh) for Horn Rev 1 has been estimated for the different scenarios with 7.7\% as the baseline ambient turbulence intensity. 
    }
    \label{tab:AEP}
    \begin{tabular}{l c c c}
 \hline\noalign{\smallskip}    
$I_u$ & HAWT (GWh)    &  VAWT (GWh) & Net (GWh)               \\
 \noalign{\smallskip}\hline\noalign{\smallskip} 
5\%   & $575.33$  & $ 103.4$ & $672.95$  \\ 
7.7\% & $577.64$  & $ 105.7$ & $677.57$   \\
15\%  & $579.95$  & $ 108.6$ & $682.77$  \\
 \hline
    \end{tabular}
\end{table}

\begin{table}
    \centering
\def~{\hphantom{0}}    
    \caption{LCoE of an offshore wind farm with and without the installation of VAWTs.  
    }
    \label{tab:LCoE}
    \begin{tabular}{l c c c c}
 \hline\noalign{\smallskip}    
$I_u$ & LCoE\textsubscript{HAWT}    &  LCoE\textsubscript{VAWT} & LCoE\textsubscript{Net} & Change in LCoE\textsubscript{Net}        \\
 & (\$/MWh)    &  (\$/MWh) &  (\$/MWh) &  (\%)               \\
 \noalign{\smallskip}\hline\noalign{\smallskip} 
5\%   & $50.8$  & $ 108.9$ & $59.65$ & $+17.4\%$ \\ 
7.7\% & $51$  & $ 110$ & $60.17$ & $+18.0\%$  \\
15\%  & $51.2$  & $ 115.5$ & $61.34$ & $+19.8\%$  \\
 \hline
    \end{tabular}
\end{table}

The expected European offshore LCoE for 2028 from \cite{Mackenzie2019} has been applied to determine the baseline for Horns Rev 1. The results indicate that even though VAWTs increase the wind farm power output, it also increases the LCoE. The LCoE for VAWTs has been estimated and downscaled using the estimations from Ref. \cite{ennis2018system}, which predicts a future LCoE of 110 \$/MWh for 5 MW floating VAWTs. However, for this example the floating foundation takes up approx 60\% of the CAPEX \cite{ennis2018system} in comparison to 22\% for the HAWTs, which is one of the reasons for the increased LCoE \cite{heptonstall2012cost}. Interesting aspects which have not been examined are the potential savings of collaborative costs for installation vessel, electrical infrastructure (approx. 19\% of combined CAPEX for offshore HAWT \cite{heptonstall2012cost}), and the increased output density MWh/km$^2$. 
\black

\section{Conclusion} \label{chp:Conclusion}
The benefits associated with co-locating HAWTs and VAWTs in a finite-size wind farm are investigated in this study. In this regard, LES together with the analytical wake model is employed. Small-scale VAWTs in triangular clusters are deployed within a finite-size wind farm consisting of conventional HAWTs. 
For the particular cases studied here, the potential power gain in the wind farm with both HAWTs and VAWTs is up to $21 \%$ compared to a baseline case in which only HAWTs are present. 
It is also shown that the impact of small-scale VAWTs on the performance of HAWTs is relatively small if the VAWTs are deployed properly among HAWTs. 
Furthermore, the performance of the analytical framework is evaluated using the LES data, and it is found that the presented analytical framework is able to accurately predict the power output from wind farms consisting of both HAWTs and VAWTs. 

The \blue calibrated \black analytical wake model is used to investigate the potential power enhancement in the Horns Rev 1 wind farm by adding small-scale VAWTs over a wide range of wind directions. It is shown that by adding the small-scale VAWTs to the wind farm, the power production can increase by up to $18\%$ (averaged over all wind directions). We conclude that co-locating conventional HAWTs and small-scale VAWTs is a promising approach to increase the land-area power density in existing wind farms. 

\blue
The LCoE for the co-located wind farm is also addressed. It is shown that having access to real CAPEX numbers for bottom-fixed foundations for VAWT would have reduced the LCoE significantly. It is further expected that floating wind turbines and VAWTs will undergo the same technological innovation patterns \cite{sovacool2015one} and reductions in LCoE  as what have been seen for offshore HAWTs in the past decades \cite{ray2019lazard} making it an interesting asset for increasing the output density for offshore wind. 

\blue
Despite the promising results presented in this study future research is required (a) to evaluate the impact of atmospheric thermal stability on the performance of co-located wind farms, (b) to extend the validation of both LES and analytical wake models to different atmospheric regimes and different wind-farm layouts, (c) to assess the effect of a realistic wind rose in the calculation of LCoE in co-located wind farms, and (d) to optimize the design and placement of clustered VAWTs within HAWT arrays to maximize the power production of co-located wind farms. 
\black

\vspace{0.2cm}
\noindent
\small{\textbf{Acknowledgments} The work is financially supported by Aarhus University.}

\section*{References}

\bibliographystyle{iopart-num}
\providecommand{\newblock}{}

\end{document}